# Signature of partially frustrated moments and a new magnetic phase in CeNiGe$_2$

Karan Singh and K. Mukherjee*

School of Basic Sciences, Indian Institute of Technology Mandi, Mandi 175005, Himachal Pradesh, India

*Email: kaustav@iitmandi.ac.in

**ABSTRACT**

We report the magnetic, thermodynamic, and transport properties of a heavy fermion compound CeNiGe$_2$. This compound undergoes two antiferromagnetic transitions around 4.1 and 3 K. It is observed in heat capacity that as magnetic field is increased to ~ 1 T, the two peak merge into a single peak around 3 K. However this peak is not suppressed under the application of magnetic field. Instead a new feature develops at 3.6 K above 1 T. The magnetic field induced new feature is investigated through entropy evolution, magnetic Gruneisen parameter and resistivity studies. These studies emphasis the fact that partial magnetic frustration due to field induced spin fluctuation is responsible for this observed feature. This partially frustrated regime develops a new antiferromagnetically ordered phase at high fields. In this compound magnetic field induced QCP is absent implying that the behavior of CeNiGe$_2$ is not in accordance to Doniach model proposed for heavy fermions compounds.

# 1. Introduction

Ternary rare-earth intermetallics compounds of the form $CeTX_2$ (T= transition metals and X= Si or Ge) have been studied extensively for many years because of their fascinating ground state, like heavy fermionic behavior in $CeRuSi_2$ and $CePtSi_2$ [1], valence fluctuation state in $CeNiSi_2$ and $CeRhSi_2$ [2, 3]. The main parameter determining such behavior in these systems depends upon the degree of localization of the 4$f$-electron on the Ce ion and the extent to which its magnetic moment is screened by conduction electrons i.e. the Kondo effect. The ground states of such Kondo systems associated with the localized 4$f$-electron has been described well by Doniach's phase diagram [4]. The Ruderman-Kittel-Kasuya-Yosida (RKKY) interaction is responsible to drive the system to an antiferromagnetically ordered state. Under external perturbations (like magnetic field, hydrostatic or chemical pressure), the magnetic ground state are tuned toward a nonmagnetic ground state in the sequence antiferromagnetic ↔ heavy fermionic ↔ intermediate valence (IV) type. These properties can be described by theoretical model arising out of competition between RKKY and Kondo interaction. The competition between these interactions also results in interestingly physical properties, like non fermi liquid (NFL) behavior, unconventional superconductivity around quantum critical point (QCP); where quantum fluctuations dominates thermal fluctuations, as observed in Ce-or Yb-based compounds [5]. Studies in such systems have also indicated a possibility of an intermediate phase induced by frustration due to additional fluctuation of local moments. Such phase forestalls the QCP and results in coexistence of screened local moment due to Kondo singlet formation and magnetically ordered moments stabilized by RKKY interactions [6].

In this context, the heavy fermion antiferromagnetic compound $CeNiGe_2$ might be interesting. Investigations carried out on single crystals of $CeNiGe_2$ under high field and external pressure [7, 8] reveals highly anisotropic magnetic and transport properties. Additionally, it is observed that these two parameters were unable to suppress the long range ordering, resulting in the absence of QCP. However, it is noted that 50% replacement of Ge by Si in this compound results magnetic order being suppressed and observation of NFL behavior [9].

In this paper, we have investigated the magnetic, thermodynamic and transport properties of the compound $CeNiGe_2$. This compound undergoes two antiferromagnetic (AFM) transitions around 4.1 and 3 K as revealed from magnetization and heat capacity studies. It is observed in heat capacity that as magnetic field is increased to ~ 1 T, the two peak merge into a single peak around 3 K. The peak is not suppressed as field is increased further; instead a new feature is observed around 3.6 K which develops into a peak as the field is increased beyond 6 T. To study the magnetic field induced new feature, we focused on the entropy evolution, magnetic Gruneisen parameter ($\Gamma_{mag}$) and resistivity studies. All these studies emphasis that magnetic frustration due to spin fluctuations is responsible for the observation of the new feature at 3.6 K (above 1 T). This partially frustrated regime develop a new ordered phase at high field (above 6 T), instead of forming the field induced QCP. Thus the behavior of $CeNiGe_2$ is not in accordance to Doniach phase diagram, being very different from that seen in other typical compounds of Ce-based HF family.

## 2. Experimental details

The compound $CeNiGe_2$ was prepared by arc melting stoichiometric amounts of respective high purity elements in an atmosphere of argon. For better homogeneity, the ingot was re-melted a number of times by turning over each time. The weight loss after the final melting was less than 1%. After melting, it was homogenized in an evacuated sealed quartz tube at $1050^0C$ for 6 days. The sample thus obtained was characterized by x-ray diffraction (XRD) and found to be single phase. X-ray diffraction pattern at room temperature was indexed to orthorhombic type CeNiSi$_2$-type structure (Figure 1(a)). The obtained lattice parameters (a = 4.260 Å, b = 16.810 Å and c = 4.209 Å) are in accordance with those reported in [2, 7]. In order to get confirmation about the stoichiometry of the compound, we performed energy-dispersive x-ray spectroscopy. The average atomic stoichiometry was found to accordance to the expected values. Temperature (*T*) (1.8 - 300 K) and magnetic field (*H*) (0 - 12 T) dependent heat capacity and resistivity were performed using Physical Property Measurement System (PPMS), while temperature dependent magnetization (*M*) were performed using Magnetic Property Measurement System (MPMS), both from Quantum design, USA. All these measurements were carried out pellets of specific shapes. For heat capacity measurement the addenda were first measured at all applied fields and was subtracted to obtain the actual heat capacity of the sample.

## 3. Results and discussion

Figure 1 (b) shows the temperature response of dc magnetic susceptibility ($\chi = M/H$) obtained under both zero-field-cooling (ZFC) and field cooled (FC) condition at 0.1 T applied magnetic field upto 10 K (upper inset shows the curve in range from 2 to 300 K). It is observed that the compound shows two antiferromagnetic (AFM) transitions around 4.1 K ($T^I_N$) and 3.2 K ($T^{II}_N$). The observed transitions are in accordance to that reported in literature for single crystals, implying the presence of two type of AFM structure in this compound [2, 7]. It is observed that there are no bifurcations in these curves which indicate to the fact that magnetocrystalline anisotropy plays a trivial role in this polycrystalline compound [10]. The inverse magnetic susceptibility is fitted with Curie-Weiss (CW) law (above 100 K) as shown in lower inset of Figure 1(b). The obtained CW temperature ($\theta_p$) is around -26 K indicating the dominance of antiferromagnetic interactions. The effective paramagnetic moment ($\mu_{eff}$) is found to be around 2.4 $\mu_B$ agree which is comparable to the theoretical value expected for the free $Ce^{3+}$ ion (2.5 $\mu_B$). This indicates that Ce moments are localized and magnetic moment in compound is due to the $Ce^{3+}$ ion sublattice only, and Ni behaves as nonmagnetic. The Kondo temperature for the compound is estimated using the formula $T_K \sim |\theta_p/2| \sim 13$ K. The deviation from CW behavior below 100 K can be attributed to crystal field effect as observed for other Ce-compounds [11]. Figure 1 (c) shows the temperature response of magnetization (*M*) under different field upto 7 T. It is observed that the *M* increases as the applied field is increased which indicates the development of ferromagnetic correlations in the antiferromagnetic state. Field response of magnetization (Figure 2 (d)) also supports this observation where a deviation (and maxima in the dM/dH curve) is observed around 0.7 T at 2 and 3 K in contrast to a smooth *M* (*H*) at 4 and 5 K.

Figure 2 (a) and (b) displays the 4*f*-electron contribution to heat capacity ($C_{4f}$) of the compound in different field in the range of 0-12 T. The 4*f*-electron contribution of heat capacity is extracted by subtracting the *C* data of YNiGe$_2$ (prepared under similar condition as that of CeNiGe$_2$) from that of CeNiGe$_2$ using the formula:

$$C_{4f}/T = C\ (CeNiGe_2)/T - C\ (YNiGe_2)/T \qquad (1)$$

At 0 T, $C_{4f}/T$ shows two peak around 3.8 and 3.1 K, which is near the $T^I_N$ and $T^{II}_N$ respectively. As the magnetic field is increased, $T^I_N$ appears to be suppressed, while $T^{II}_N$ remains unchanged. Around 1 T only one peak is observed around 3 K. As the magnitude of field increases (beyond 3 T) it results in a broad peak around $T^{II}_N$. Interesting, another weak anomaly is observed in the curve around 3.6 K (above 1 T) which become a pronounced peak as the field is increased to 6 T. Interestingly, this peak temperature is shifted downward in temperature as the field is increased. Such a feature, that is, the decrease of the peak temperature under an application of magnetic-field, is characteristic of antiferromagnetic ordering. Thus in this compound a new magnetic ordering is developed around 3.6 K beyond 6 T. It implies that due to appearance of this new feature this compound avoids field induced QCP. Similar type of behavior is also observed in this compound under the influence of external pressure [8]. In low temperature region, we have calculated the linear Somerfield coefficient ($\gamma$) using the formula $C_{4f}/T = \gamma + \beta T^2$. It is observed that $\gamma = 433$ mJ/mol K$^2$ at 0 T which is comparable to the $\gamma$ value reported in [9]. This high value of $\gamma$ confirms CeNiGe$_2$ is a heavy fermions compound. The value of $T_K$ determined from $\gamma$ [12], is found out to be ~ 13 K, which is in accordance to that estimated from CW temperature.

To get further information about the field induced new feature, field dependence of $C_{4f}/T$ is plotted at selected temperatures near the observed magnetic transitions (as shown in Figure 3 (a)). In the temperature range of 2.4 to 3.2 K, $C_{4f}/T$ increases with decreasing field and saturates below 2 T. At 3.6 K a broad peak is observed while such features are absent at temperatures above 4 K. Hence it can be said that variation of applied magnetic field induces the new features in $C_{4f}$ which changes as the temperature is increased. A sharp kink in the field response of $C_{4f}$ is noted for systems which favors long range magnetic ordering. In our case the observation of the broad feature can be ascribed to presence ferromagnetic correlation in antiferromagnetic state. This competition between ferromagnetic and antiferromagnetic interactions results in spin fluctuations in this compound in accordance to that observed in Ref [13]. Therefore it can be said that in this compound above 3.2 K, field induced spin fluctuation develops which coexists with magnetic ordering. The presence of spin fluctuation in a magnetically ordered state has also been observed for other polycrystalline compounds [13, 14]. Hence the weak anomaly is observed in the $C_{4f}/T$ vs. $T$ curve around 3.6 K (at 1T) can be ascribed to magnetic frustration which arises due to field induced spin fluctuations. In order to probe the magnetic low-energy states of this compound, we have studied the field dependence of magnetic entropy ($S_{4f}$). Inset of Figure 2 (b) displays the $T$ dependence of $S_{4f}$. Around $T_K$ the observed entropy is around 0.52 Rln2. This value further reduces to ~ 0.22 Rln2 and 0.13 Rln2 at $T^I_N$ and $T^{II}_N$ respectively. This low value of the entropy implies a partial screening of Ce magnetic moment by conduction electron spins. Hence this result implies a coexistence of the Kondo effect with magnetic ordering which is in accordance to that observed for other Ce-compound [15]. Figure 3 (b) shows the field dependence of $S_{4f}$ at selected temperatures. The entropy shows a maximum around 2.5 T at 2 K which at 3.6 K reduces to ~ 1.25 T. Above the magnetic ordering temperature around 5 K, no such maxima is observed and entropy decreases with increasing field. Such decrease of magnetic entropy is related to the suppression of magnetic correlations due to magnetic field.

Further to substantiate the claim about the role magnetic frustration in this compound, we have calculated the magnetic Gruneisen parameter ($\Gamma_{mag}$). In these types of compounds, $\Gamma_{mag}$ is an excellent tool to indentify the supposed magnetic frustration [16]. Generally, $\Gamma_{mag}$ displays a sign change in the frustration regime, signaling entropy accumulate at this regime [17]. The $\Gamma_{mag}$ is calculated, using the formula

$$\Gamma_{mag} = - (dS/dH)/C \qquad (2)$$

where $dS/dH$ is the field derivative of the entropy [17]. $H$ response of $\Gamma_{mag}$ at different temperatures is plotted in Figure 4(a). A crossover from a positive to a negative value due to entropy accumulation is noted around fields which are nearly equal to the field where a maxima is observed in field dependence of $S_{4f}$. At 3.6 K, the crossover is observed around 1 T, which is the same field where a weak anomaly is observed in the temperature response of $C_{4f}$. Hence the field around 1 T can be considered as the critical field ($H_C$) of this compound. Also at this temperature a crossover is observed implying that the compound is frustrated around this temperature and is undecided which ground state to be chosen. This behavior is in accordance to that observed Ni substituted CePdAl [18]. Figure 4(b) shows the $T$ dependence of $\Gamma_{mag}$ in field range 1 - 6 T. The fact that the data at 1 T display a zero crossing around 3.6 K is consistent with heat capacity data analysis.

In this compound, as shown in Figure 5, the data at various field (upto 7 T) collapse on a single curve when plotted as $\Gamma_{mag} h$ versus $T/h^\varepsilon$ where $h = H - H_C$ and $\varepsilon$ is scaling exponent [19, 20]. The best fit is obtained for $\varepsilon = 0.35$ and $H_C = 1.5$ T is consistent with the critical field obtained from the heat capacity analysis. The excellent collapse of the data demonstrate that CeNiGe$_2$ have a frustrated regime at low field. We have observed a crossover energy scale at 2.5 K/$T^{0.35}$. The curve above and below 2.5 K/$T^{0.35}$, is best fitted with $x^{-1.3}$ and $x^{3.5}$ (where $x = T/h^{0.35}$) respectively. Hence, $\Gamma_{mag} \sim T^{-1} h^{-0.4}$ and $\sim T^{-0.7} h^{0.4}$ above and below 2.5 K/$T^{0.35}$ respectively. Generally for spin fluctuations in ordered state, $\Gamma_{mag}$ varies as $T^{-\alpha}$, where $\alpha$ is temperature exponent [21, 22]. In our compound below 2.5 K/$T^{0.35}$, $\alpha$ is nearly equal to 2/3. This indicate that the observed crossover in variation of $\Gamma_{mag}$ in this compound can be ascribed to the change over from paramagnetic regime (> 3.6 K) to field induced spin fluctuations regime (around 3.6 K). This partially frustrated regime develop a new ordered phase at high field (above 6 T), instead of forming the field induced QCP.

Figure 6 (a) shows the temperature dependence electrical resistivity ($\rho$) in field range 0 – 7 T in the temperature range of 0 to 50 K. As the temperature is reduced, $\rho$ decreases. At 0 T, around 13 K, a minima in resistivity is seen which is ascribed to Kondo effect. This observation is consistent with $T_K$ obtained from magnetization and heat capacity data. This feature is suppressed as the magnitude of field increases. Inset of the Figure 6 (a) show the temperature dependence of first derivative of resistivity (d$\rho$/dT). At low fields two slope changes around $T^I_N$ and $T^{II}_N$ are observed. As the magnitude of field is increased beyond 6 T, the slope changes are replaced by a broad hump around 3.6 K. Hence there is a good agreement between the features observed from magnetization, heat capacity and $\rho$ measurements. Field response of first derivative of resistivity at selected temperatures near the magnetic transitions is presented in Figure 5 (b). A minima is observed and its value increased from the 1.8 to 3.6 K. Above 3.6 K, the value of the minima decrease. This observation gives further evidences to the fact that around 3.6 K, spin-fluctuations arises. The results of $\rho$ measurements indicates that observed features in this compound at 3.6 K is due to the magnetic frustration arising from the spin fluctuations due to Ce moments which is also in analogy from our observation from the thermodynamic measurements.

## 4. Summary

In summary, we report the magnetic, thermodynamic, and transport properties of a heavy fermion compound $CeNiGe_2$. This compound undergoes two antiferromagnetic transitions around 4.1 and 3 K. From heat capacity studies, it is observed that around 1 T the two peaks merge into a single peak around 3 K. As the magnetic field is increased further the peak is not suppressed; instead a new feature is seen around 3.6 K above 1 T. The magnetic field induced new feature is investigated through entropy evolution, magnetic Gruneisen parameter ($\Gamma_{mag}$) and resistivity studies. Our results point to the fact that magnetic frustration due to spin fluctuations is responsible for the observed feature at 3.6 K (above 1 T). This partially frustrated regime develops a new antiferromagnetically ordered phase at high fields. Further experimental probes like elastic neutron diffraction are necessary to explore this new phase. In this compound magnetic field induced QCP is absent. Hence the behavior of $CeNiGe_2$ is not in accordance to Doniach model which is widely used to classify heavy-fermion compounds.

**Acknowledgements**

The authors acknowledge experimental facilities of Advanced Material Research Centre (AMRC), IIT Mandi. Financial support from IIT Mandi is also acknowledged.

**Figures**

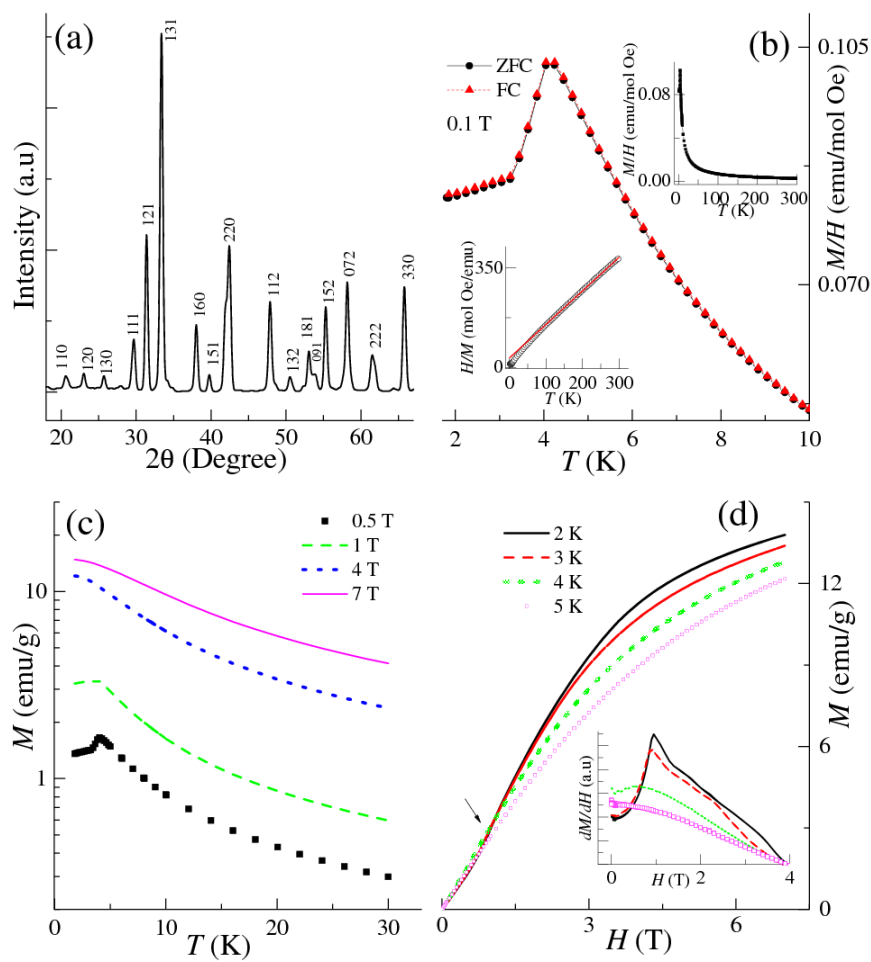

Figure 1: (a) X-ray diffraction patterns for the compound. (b) Temperature (*T*) response of the dc magnetic susceptibility in field 0.1 T upto 10 K under both ZFC (Zero field cooled) and FC (field cooled) condition. Upper inset: Same plot in the *T* range of 1.8 – 300 K. Lower inset: *T* dependence of inverse magnetic susceptibility in at 0.1 T. Solid red line is the Curie-Weiss law fitting. (c) *T* (1.8 – 30 K) response of magnetization at fields 0.5, 1, 4, and 7 T. (d) Isothermal magnetization (*M*) vs. applied field (*H*) curves at temperatures 2, 3, 4 and 5 K. Inset: *dM/dH* vs. *H* plots at same temperatures.

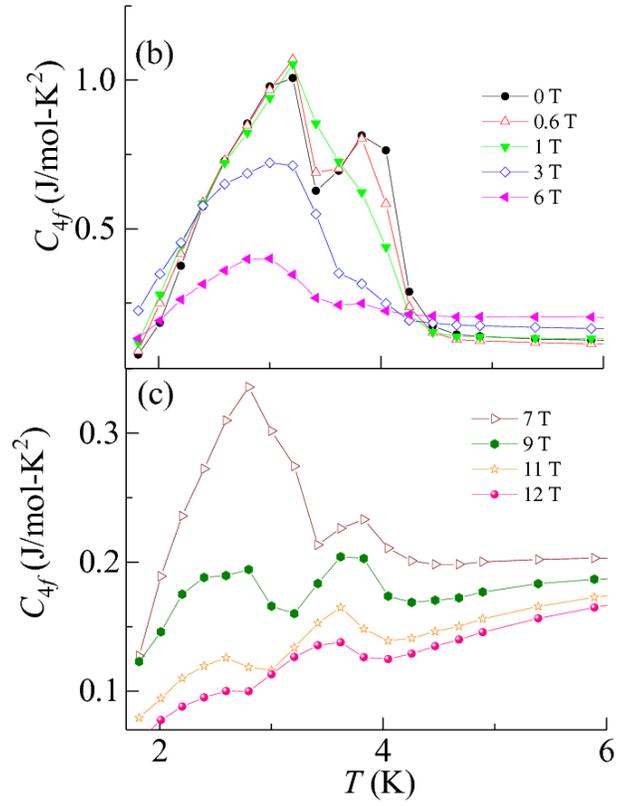

Figure 2: (a) Temperature (*T*) response of 4*f*-electron contribution to the heat capacity ($C_{4f}$) in the field range of 0 – 6 T. (b) Same plot in the field range of 7 - 12 T.

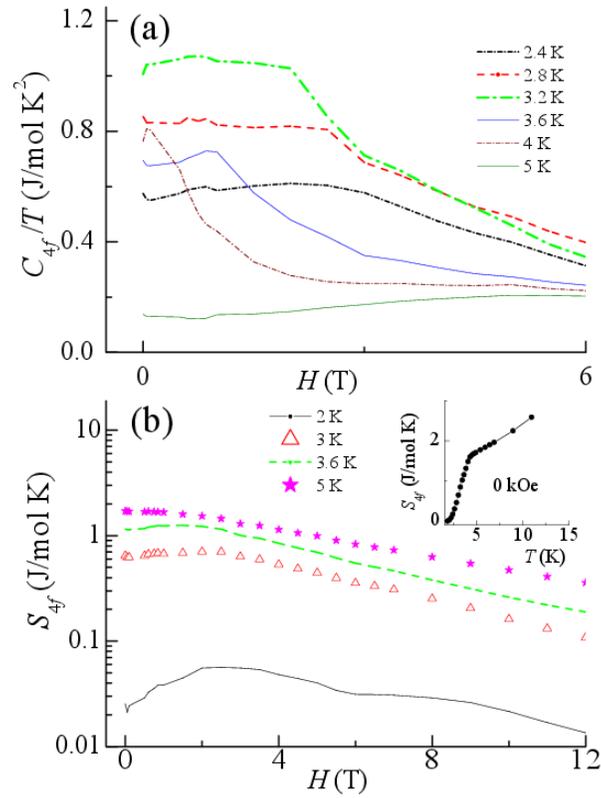

Figure 3: (a) Magnetic field response of magnetic entropy ($S_{4f}$) at selected temperatures (2.4, 2.8, 3.2, 3.6, 4 and 5 K) upto 6 T. (b) $S_{4f}$ as a function of magnetic field at 2, 3, 3.6 and 5 K. Inset: Temperature response of $S_{4f}$ in zero field.

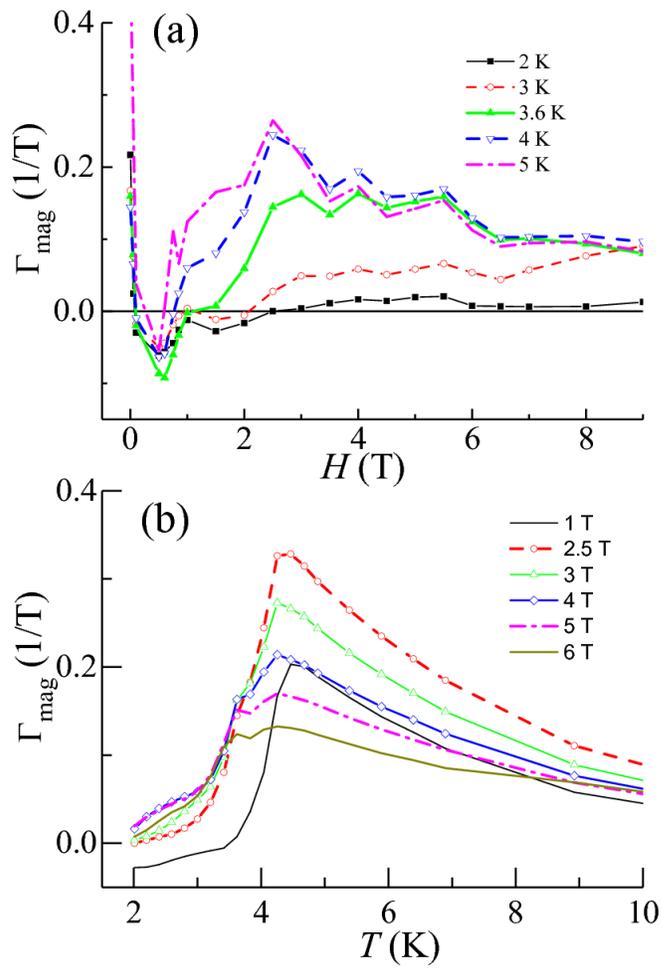

Figure 4: (a) Field-dependence of the magnetic Gruneisen parameter ($\Gamma_{mag}$), at temperature 2, 3, 3.6, 4, and 5. (b) Temperature response of $\Gamma_{mag}$ in field range of 1 - 6 T.

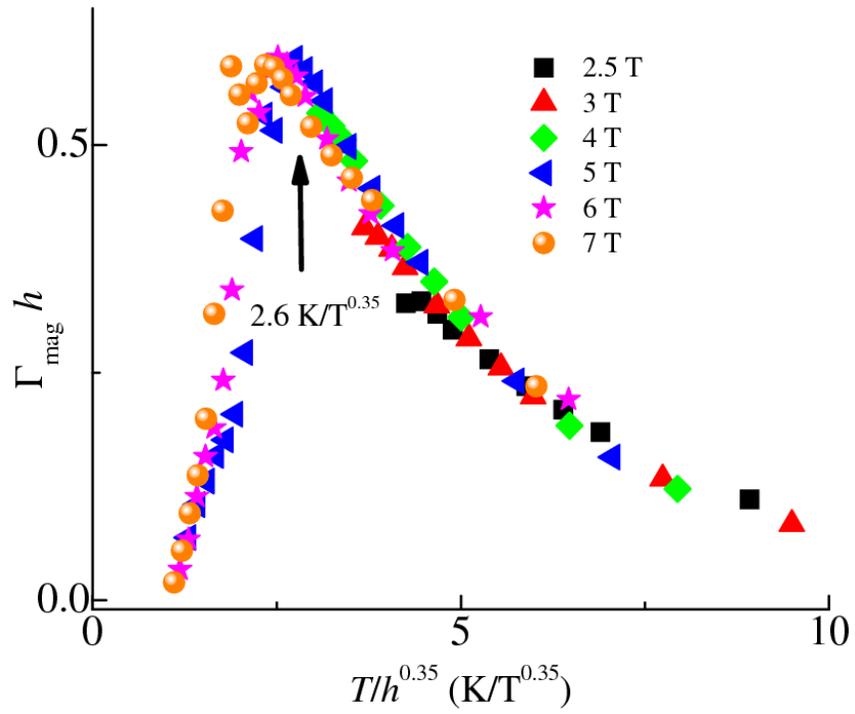

Figure 5: Scaling plot of $\Gamma_{mag} h$ vs $T/h^\varepsilon$ with $\varepsilon = 0.35$ and $h = (H - 1.5)$ T

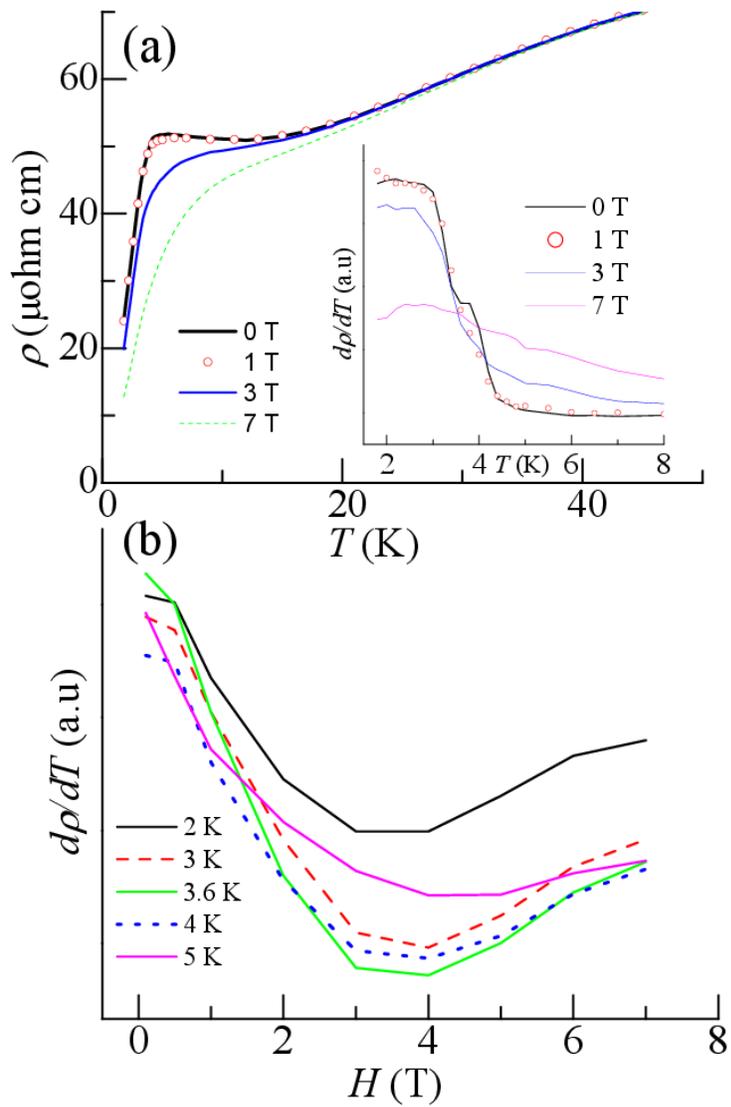

Figure 6: (a) Resistivity (ρ) plotted as a function of temperature (*T*) in the field range of 0 - 7 T upto 50 K. Inset: *T* response of temperature derivative of ρ upto 8 K in the field range of 0 - 7 T. (b) *H* response of temperature derivative of ρ at 2, 3, 3.6, 4 and 5 K.

17